\documentclass[aps,twocolumn,showpacs,floatfix]{revtex4}
\usepackage{amsmath,amsfonts,amssymb,graphicx,color,times,bbm,psfrag,dsfont}
\usepackage[T1]{fontenc}

\newcommand{\bra}[1] {\langle{#1}\vert }
\newcommand{\ket}[1] {\vert{#1}\rangle }
\begin{document}

\title{Quantum algorithm for the Laughlin wave function}

\author{J. I. Latorre, V. Pic\'o and A. Riera}

\affiliation{
Dept. d'Estructura i Constituents de la Mat\`eria,
Universitat de Barcelona, 647 Diagonal, 08028 Barcelona, Spain
}
%\date{\today}

\begin{abstract}
We construct a quantum algorithm that creates the Laughlin state for an arbitrary
number of particles $n$ in the case of filling fraction one. This 
quantum circuit is efficient since it only uses  $n(n-1)/2$ local qudit gates and its 
depth scales as $2n-3$. 
We further prove the optimality of the circuit using permutation theory arguments and 
we compute exactly how entanglement develops along the action of each gate. 
Finally, we discuss its experimental feasibility decomposing the qudits and the gates in terms
of qubits and two qubit-gates as well as the generalization to arbitrary filling fraction.
\end{abstract}
\maketitle

One of the most important goals in the field of 
quantum computation is to achieve a faithful and efficient simulation 
of relevant quantum systems. Feynman first suggested
\cite{F82,NC00} the possibility of emulating
a quantum system by means of other specially designed 
and controlled quantum systems.
Nowadays,  ultra-cold atoms in optical lattices  
are producing the first quantum simulators 
\cite{LSADSS07-BDZ08}
and many theoretical proposals have been already presented in order
to approximately simulate strongly correlated systems \cite{PPC,HSDL07,BHHFZ05,PB08,MBZ06,DDL03}.

A more ambitious goal consists in finding the {\sl exact quantum circuit}
that underlies the physics of a given quantum system. 
Rather than searching for an analogical simulation, 
such an exact quantum circuit would fully reproduce the properties of
the system under investigation without any approximation and 
for any experimental value of the parameters in the theory.
It is particularly interesting to device new
quantum algorithms for strongly correlated quantum systems of few
particles. These could become the first non-trivial uses of
a small size quantum computer. It is worth recalling that 
strong correlations are tantamount to
a large amount  of entanglement in the quantum state and this,
in turn, implies that the system is
hard to be simulated numerically \cite{G03}.

An exact quantum circuit
 would start from a simple product unentangled
state and would create faithfully any desired  
state  on demand. That is, such a quantum algorithm would
diagonalize the dynamics of the target system. At present, very few cases are under
control \cite{BS08,VCL08}. In Ref. \cite{VCL08}, the underlying quantum circuit
that reproduces the physics of the thoroughly studied XY Hamiltonian was obtained.
The philosophy inspiring that circuit was to follow the
steps of the analytical solution of that integrable model.
It is not obvious how to design a quantum simulator for non-integrable systems.
Here, we shall present a quantum circuit that allows the controlled
construction of a particular case of the Laughlin wave function.
Thus, the quantum algorithm we are putting forward will
not produce the complete dynamics of a Hamiltonian but rather
a specific state. That is, our quantum circuit will transform a trivial
product state into a Laughlin state with filling fraction one by means of a finite
amount of local two-body quantum gates.

Let us start by recalling the Laughlin \cite{L83}
wave function for filling fraction $\nu=1/m$, which corresponds to
\begin{equation}
\label{eq:Laughlin_state}
\Psi_L^{(m)}(z_1,\ldots,z_n)\sim
\prod_{i<j}(z_i-z_j)^{m} 
e^{-\sum_{i=1}^{N}\vert z_i \vert^2/2} \ ,
\end{equation}
where $z_j=x_j+\textrm{i}y_j, j=1,\ldots,n$ 
stands for the position of the $j$-th particle.
This state was postulated by Laughlin as
the ground state of the fractional quantum Hall effect (FQHE) \cite{Tsui82}. 
From the quantum information point of view, 
the Laughlin state exhibits a considerable von Neumann entropy 
between for any of its possible partitions \cite{ILO07}.
It is, thus, classically hard to simulate such a wave function
making it an ideal problem for a quantum computer.

We shall construct the Laughlin state using a
quantum system that consists of a chain of $n$ {\sl qudits} 
($d$-dimensional Hilbert spaces). 
In our case, that is $m=1$, the dimension of the qudits is needed to be $d=n$.
Let us proceed to construct the quantum circuit by first considering
the case of $n=2$ particles, then $n=3$ and, finally, the general case.

The Laughlin state can be written in terms of the single particle
angular momentum eigenstates, also called Fock-Darwin states
$\varphi_{l}(z)=\bra{z}l\rangle=z^l \exp(-|z|^2/2) / \sqrt{\pi l!}$.
Then, the $n=2$ Laughlin state reads 
\begin{equation}
  \Psi_L^{(2)}(z_1,z_2)=
  \frac{1}{\sqrt{2}}\left(\varphi_1(z_1)\varphi_0(z_2)-\varphi_0(z_1)\varphi_1(z_2)\right) \ .
\end{equation}
Let us note that $ \Psi_L^{(2)}(z_1,z_2) =\bra{z_1,z_2}\Psi_2^{(2)}\rangle$
is simply the projection of the Laughlin state
\begin{equation}
  \ket{\Psi_L^{(2)}}=\frac{1}{\sqrt{2}} \left(\ket{0,1}-\ket{1,0}\right)
	\, 
\end{equation}
in coordinates representation, 
where particle label is retained in the order of qubits and
the angular momentum 0 or 1 is an element of the angular momentum basis.

\begin{figure}
\scalebox{0.8}{\includegraphics{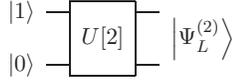}}
\caption{Scheme for the quantum circuit that generates the $n=2$ Laughlin state
by acting on the product state $ \vert \Psi_0^{(2)} \rangle=\ket{0,1}$.
Notice that the qubits of the input state are labeled from bottom to top.}
\label{fig:two-qubit-gate}
\end{figure}

It is trivial to find a quantum circuit that
transforms a product state into the above $n=2$ Laughlin state. 
Let us first prepare an initial state as 
$\ket{\Psi_0^{(2)}}=\ket{0,1}$ and  perform on it the simple two-qubit gate $U[2]$
as shown in Fig. \ref{fig:two-qubit-gate}. The exact form of 
the unitary operator $U[2]$ in the angular momentum basis
$ \{ \ket{0,0},\ket{0,1},\ket{1,0},\ket{1,1} \}$ is
\begin{equation}
U[2]=
\begin{pmatrix}
1 & 0 & 0 & 0 \\
0 & \frac{1}{\sqrt{2}} & \frac{1}{\sqrt{2}} & 0 \\
0 & -\frac{1}{\sqrt{2}} & \frac{1}{\sqrt{2}} & 0 \\
0 & 0 & 0 & 1  
\end{pmatrix}
\, .
\label{eq:n2gate}
\end{equation}

Let us now move to the more complicate case of the
Laughlin state with three
particles, $n=3$. We now need to consider a
system of three qutrits $d=n=3$.
Following similar steps as we did for $n=2$, 
we take as initial state $\ket{\Psi_0^{(3)}}=\ket{0,1,2}$,
that is, each qutrit is prepared in a different basis element,
representing different angular momenta.
The aim of the quantum circuit is to antisymmetrize this initial state,
since the Laughlin wave function for $m=1$ is simply the Slater determinant
of the single particle wave functions 
\begin{align}
\Psi_L^{(3)}(z_1,z_2,z_3)&=\bra{z_1,z_2,z_3}\Psi_L^{(3)}\rangle \nonumber \\
&=\frac{1}{\sqrt{6}} 
\left\vert
\begin{array}{ccc}
 \varphi_0(z_1) & \varphi_0(z_2) & \varphi_0(z_3) \\
 \varphi_1(z_1) & \varphi_1(z_2) & \varphi_1(z_3) \\
 \varphi_2(z_1) & \varphi_2(z_2) & \varphi_2(z_3) 
\end{array}
\right\vert
 \, .
\label{eq:n3Laughlin}
\end{align}
To do this, we define the two-qutrit unitary operators $W_{ij}(p)$ as
\begin{align}
W_{ij}(p)\ket{ij}&=\sqrt{p} \ket{ij}-\sqrt{1-p} \ket{ji}  \nonumber \\
W_{ij}(p)\ket{ji}&=\sqrt{p} \ket{ji}+\sqrt{1-p} \ket{ij}   \, ,
\label{eq:W-gates}
\end{align}
for $i<j$, $0 \le p\le 1$, 
and  $W_{ij}\ket{kl}=\ket{kl}$ if $(k,l)\ne (i,j)$. 
We realize that for the case of qubits and $p=1/2$ we recover the 
gate $U[2]$ of Eq. (\ref{eq:n2gate}).
Let us note that the unitary operator $W_{ij}$ is
a linear combination of the identity ($p=1$) 
and the {\sl simple transposition} ($p=0$) operators, where 
 a simple transposition is defined as the 
transposition between two contiguous elements.
\begin{figure}
\scalebox{0.8}{\includegraphics{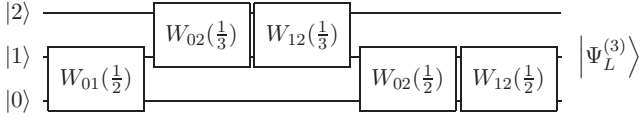}}
\caption{Quantum circuit that produces the $n=3$ Laughlin state acting on 
a product state $ \vert \Psi_0^{(3)} \rangle=\ket{0,1,2}$.}
\label{fig:n3circuit}
\end{figure}
The architecture of the quantum circuit that produces 
the $n=3$ Laughlin state in Eq. (\ref{eq:n3Laughlin}) by means of
the local gates from Eq. (\ref{eq:W-gates})
is presented in Fig. \ref{fig:n3circuit}.

So far, we have seen the quantum circuits that produce the Laughlin
state for $n=2$ and 3. From these cases, a general
scheme emerges that will produce the correct quantum circuit
for an arbitrary number of particles. Let us proceed by induction. 
We will
assume that we already know the quantum circuit, $U[n]$, that produces
the Laughlin state for $n$ qudits when acting
on $\ket{\Psi_0^{(n)}}=\ket{0,1,2,\dots,n-1}$. We now need
to complete the circuit to achieve the $n+1$ Laughlin state
from the product state $\ket{\Psi_0^{(n+1)}}=\ket{0,1,\dots,n-1,n}$. 

The Laughlin state for $n$ qudits has the form
\begin{equation}
  \ket{\Psi_L^{(n)}}=\frac{1}{\sqrt{n!}}
  \sum_{\mathcal{P}} \textrm{sign} (\mathcal{P}) 
  \ket{a_1, \ldots, a_n} \, ,
 \label{eq:Laughlin-expanded}
\end{equation}
where the sum runs over all the possible permutations 
of the set $\{0,1,\ldots,n-1 \}$ and, given a permutation,
$a_k$ represents its $k$-th element.
The relative sign between two permutations corresponds 
to the parity of the number of transpositions needed to
transform one into the other.
If we add another qudit, the system is in a product state
$\ket{\Psi_L^{(n)}} \ket{n}$.
According to Eq. (\ref{eq:Laughlin-expanded}), 
we want to generate a superposition of $(n+1)!$ permutations
corresponding to the $n+1$ states of the Laughlin wave function, 
from the superposition of $n!$ permutations that we already have
in the $n$-qudit case.

Let us note that, if we have the set of $n!$ permutations of $n$ 
elements, 
we can generate the set of permutations of $n+1$ elements 
by performing successive simple transpositions between
the new element and its preceding neighbour in the sequence
$\left( \{a_1,\ldots,a_{n},n\}\right.$, 
$\{a_1,\ldots,n,a_{n}\}$,
$\ldots$, $\left. \{n,a_1,\ldots,a_{n}\}\right)$. 
\begin{figure}
\scalebox{0.8}{\includegraphics{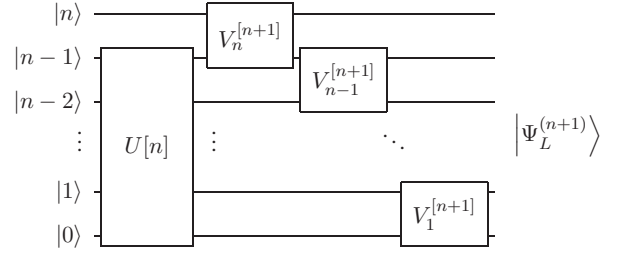}}
\caption{Quantum circuit that produces the Laughlin state for an arbitrary number
of wires $(n+1)$ acting on a product state $\ket{\Psi_0^{(n)}}=\ket{0,1,2,\cdots, n}$.
Note its recursive structure,
$U[n+1]=V_1^{[n+1]} V_2^{[n+1]} \ldots V_n^{[n+1]} U[n]$,
where $V$ gates are defined in Eq. (\ref{eq:V-gates}).
}
\label{fig:recursive-circuit}
\end{figure}
This idea suggests a circuit as the one shown 
in Fig. \ref{fig:recursive-circuit}. 
In this scheme, the gate $V_n^{[n+1]}$ should produce a superposition of
all the permutations $\ket{a_1, \ldots ,a_{n},n}$ 
with $\ket{ a_1, \ldots, n ,a_{n} }$. 
The gate $V_{n-1}^{[n+1]}$ should do the same task in the next site, that is, should produce
a superposition between
$\ket{a_1, \ldots ,a_{n-1},n,a_{n}}$
and $\ket{a_1, \ldots ,n,a_{n-1},a_{n}}$. This scheme works
successively till $V_1^{[n+1]}$.

This general structure implies that $V_n^{[n+1]}$ has to be decomposed in terms of the 
$W$-gates presented previously, 
\begin{equation}
V_n^{[n+1]} = W_{0n}(p)W_{1n}(p)\ldots W_{n-1\,n}(p) \, , 
\end{equation}
where $p$ is a common weight due to the fact that the 
$0, \ldots, n-1$ states in the Laughlin wave function are indistinguishable.
Let us note that all the operators $W_{in}$ in the previous expression
commute among themselves and, therefore, the order in which they are applied
is irrelevant.

In order to determine the weight $p$ in the $V_n^{[n+1]}$ gate, 
we realize that if all the transpositions only involve the state $n$, 
the states $\ket{a_1, \ldots,a_{n},n}$ will not
be affected by the rest of gates, and they should already have the correct normalization
factor $\frac{1}{\sqrt{(n+1)!}}$ after applying $V_n^{[n+1]}$. This implies $p=\frac{1}{n+1}$.

We proceed in a similar way to determine the rest of gates, and we obtain
\begin{equation}
V_k^{[n+1]}=\prod_{i=0}^{n-1} W_{in}\, ,
\label{eq:V-gates}
\end{equation}
where, given a $k$, all the $W$ gates have the same 
weight $\frac{1}{k+1}$ and $k=1\ldots n-1$. 
Notice that the $W_{ij}$ gates 
act on states with $i<j$ along the whole circuit and, therefore, 
they always generate the negative combination 
$\sqrt{p}\ket{ij}-\sqrt{1-p}\ket{ji}$. 
This is the reason why each term of the final state has 
the appropriate sign, 
since a minus sign is carried in each transposition. 

\begin{figure}
\scalebox{0.8}{\includegraphics{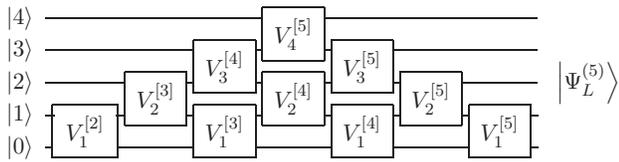}}
\caption{Quantum circuit that produces the Laughlin state of 5 particles acting on a product state $\ket{01234}$.}
\label{fig:qcircuit-n5}
\end{figure}

The above discussion produces our main result, that is, the
circuit shown in
 Fig. \ref{fig:recursive-circuit} that uses the definition of 
its gates in Eq. (\ref{eq:V-gates}). Such a quantum circuit will 
generate the $m=1$ Laughlin wave function
for an arbitrary number of qudits. 
In particular, the quantum circuit corresponding to 5 qudits 
is presented in Fig. \ref{fig:qcircuit-n5}. 

%\section{Circuit analysis}

%\subsection{Scaling of the number of gates and the depth of the circuit respect to $n$}
The recursive structure of the circuit (Fig. \ref{fig:recursive-circuit})
makes easy to calculate how the number of gates $N(n)$ and the 
depth $D(n)$ of the circuit scale with the total number of particles $n$. 
An elementary counting gives the result 
\begin{align}
N(n)&=(n-1)+N(n-1)=\frac{n(n-1)}{2} \nonumber \\
D(n)&=D(n-1)+2= 2n-3, 
\end{align}
with $N(2)=1$ and $D(2)=1$. 
The quantum circuit that delivers the Laughlin state is, thus,
efficient since the number of gates scales polynomially.
This is a non-trivial result since, in general, an arbitrary unitary transformation
requires an exponential number of gates to be performed.

%\subsection{Proof that the circuit is minimal}
Let us now discuss the optimality of our quantum circuit.
As we mentioned previously, a simple transposition, 
$s_i$, is defined as the transposition
between two contiguous elements, $i$ and $i+1$.
Any permutation can be decomposed in terms of a 
series of simple transpositions
and its minimal decomposition is called the 
{\it canonical reduced decomposition}.
There are two particular interesting permutations: 
({\it i}) the minimum permutation ($0,1,2,\ldots,n-1$) 
whose canonical reduced decomposition is the identity, 
and ({\it ii)} the maximum permutation ($n-1,n-2, \ldots,0$)
whose canonical reduced decomposition reads
%\begin{equation}
$s_1(s_2 s_1)(s_3 s_2 s_1)\ldots (s_{n-1} \ldots s_1) \, ,$
%\end{equation}
and it is the permutation with the largest number of 
simple transpositions in its canonical reduced decomposition.
Then, a circuit that produces the state
corresponding to the maximum permutation needs as many
gates as the number of simple transpositions of its canonical
reduced decomposition, that is $\frac{n(n-1)}{2}$.
The Laughlin state contains this maximum permutation, therefore,
its quantum circuit must have, at least, $\frac{n(n-1)}{2}$ gates,
that is precisely the number of gates in our proposal.

%\subsection{Entanglement}
It is also possible to analyze the way entanglement grows along
the circuit. In order to do this, we calculate how much entanglement each gate
of the circuit generates, that is, we determine the increase of the von Neumann entropy between
the two parts of the system separated by a given gate,
\begin{equation}
\Delta S \left( V_k^{[n]} \right)=\log_2\left(\frac{n}{n-k} \right) \, .
\end{equation}
The von Neumann entropy between $k$ particles and the rest of the system is
simply the sum of the contributions of those gates that are in the row
which separates the system in $k$ and $n-k$ wires.
These gates are $V_k^{[n']}$ for $k+1 \le n'\le n$ and the entanglement entropy reads
\begin{equation}
  S_{n,k} = \sum_{n'=k+1}^n \Delta S \left( V_k^{[n']} \right) 
          = \log_2\binom{n}{k}\, .
\label{eq:entanglement-entropy}
\end{equation}
This expression recovers in a clean way the same result as the one 
found in Ref. \cite{ILO07}, which was proven exact. 
Let us also remark that although each single particle is 
maximally entangled with the rest of the system, a subset of $k\le n/2$ particles does not saturate
the entropy, 
\begin{equation}
S_{n,k}\le k\log_2{n} \, .
\end{equation}

%\section{Experimental realization}
A experimental realization of our proposal will probably need
to work on qubits rather than qudits. It is, then, necessary to
find an efficient reduction of our algorithm to qubits. 
The easiest way to encode a qudit in terms of qubits is the
binary basis. Then, an arbitrary single state $\ket{i}$
can be decomposed as
%\begin{equation}
$\ket{i}=\ket{i_r}\ldots \ket{i_2}\ket{i_1} \, ,$
%\end{equation}
where $i=\sum_{k=1}^{r} 2^{k-1}i_k$, $r\sim \log_2 n$ is the number 
of bits needed to represent $n$, and $i_k=0,1$ $\forall k=1,\ldots,r$.

Now, we want to find the gates that acting on these qubits
implement the $W$ gates. $W_{ij}$ acts non trivially on the 
space spanned by the computational basis states 
$\ket{i j}\equiv\ket{i_r, \ldots, i_1,j_r , \ldots, j_1}$ 
and
$\ket{j i}\equiv\ket{j_r, \ldots, j_1, i_r , \ldots, i_1}$,
and is the identity for the rest of states. 
Let us define $\tilde{W}$ as the non trivial $2 \times 2$
sub matrix of $W$ that acts on this subspace.
According to Eq. (\ref{eq:W-gates}), $\tilde{W}$ takes the form
\begin{equation}
\tilde{W}\equiv
\begin{pmatrix}
 \sqrt{p} & \sqrt{1-p} \\
 -\sqrt{1-p} & \sqrt{p} 
\end{pmatrix}
\, ,
\end{equation}
and it corresponds to the exponentiation 
$\tilde{W}=\exp{\left(\textrm{i} \theta \sigma_y \right)}$ of
the $\sigma_y$ Pauli matrix for $p=\cos^2 \frac{\theta}{2}$.
In order to implement an arbitrary $W_{ij}$ gate, 
we have to follow three steps:
({\it i}) first, we compare the binary expressions of 
$i_r \ldots i_1 j_r \ldots j_1$ and 
$j_r \ldots j_1 i_r \ldots i_1$, and
notice which bits are different.
Then, we carry out a sequence of binary numbers, starting with $ij$
and concluding with $ji$, such that adjacent members of the list differ in
only one bit. These sequences are called {\sl Gray codes} \cite{NC00}. 
({\it ii}) Next, we implement a quantum circuit performing a series of 
multi-qubit controlled gates that change the state $ij$
according to the previous sequence. Each multi-qubit gate
transforms the corresponding state of the sequence into the next one.
These multi-qubits gates are carried out
until it only remains a different bit between the last transformed state and $ji$.
({\it iii}) At this point, we perform a controlled-$\tilde{W}$ gate,
or alternatively its complex conjugate $\tilde{W}^\dagger$, 
taking this different qubit as target. 
We will apply $\tilde{W}$ or $\tilde{W}^\dagger$ depending on the
initial state in which the single gate is performed in order
to tune the behaviour of $W$ according to its definition in Eq. (\ref{eq:W-gates}).
Finally, the reversed previous sequence 
of multi-qubit controlled gates is performed.

\begin{figure}
\scalebox{0.9}{\includegraphics{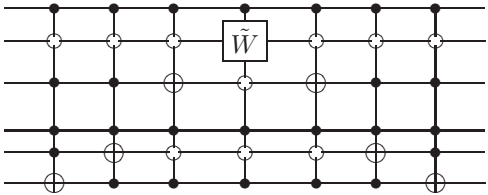}}
\caption{Implementation of the $W_{3 5}$ gate in terms of multi-qubit control gates. 
Each of these controlled operations can be decomposed in single qubit and CNOT gates.}
\label{fig:w-35}
\end{figure}

This abstract construction can be illustrated with the example of the $W_{35}$ gate 
that acts non-trivially on the states 
$\ket{011 \ 101}$ and $\ket{101 \ 011}$.
One possible sequence of Gray codes that connect $011 \ 101$ and $101 \ 011$
is
\begin{equation}
\begin{array}{ccccccc}
0       & 1       & 1 &\ \  & 1       & 0 & 1 \\ 
{\bf 1} & 1       & 1 &\ \  & 1       & 0 & 1 \\
1       & {\bf 0} & 1 &\ \  & 1       & 0 & 1 \\
1       & 0       & 1 &\ \  & {\bf 0} & 0 & 1 \\
1       & 0       & 1 &\ \  & 0       & 1 & 1 \, .
\end{array}
\end{equation}
From this, we can read its corresponding circuit, 
shown in Fig. \ref{fig:w-35}. 
Notice that the first three multi-qubit controlled gates
transforms $\ket{011 \ 101}$ into $\ket{101 \ 001}$.
Next, the $\tilde{W}$ gate is applied to the fifth qubit,
affecting only 
the states $\ket{101 \ 001}$ and $\ket{101 \ 011}$ due to the conditions
on rest of qubits. Finally, we reverse the application of 
the multi-qubit controlled gates, ensuring that $\ket{101 \ 001}$
gets swapped back with $\ket{011 \ 101}$.
It is important to point out that these multi-qubit controlled gates are not
two-qubit gates, as, in principle, it would be suited. 
Nevertheless, it is known that these controlled operations can be performed
by means of $O(r)$ single qubit and CNOT gates \cite{MMKIW95, BBCDMSSSW95}
which can be implemented experimentally \cite{CZ95, CCEHRSZ06}. 
Thus, if we consider that the number of $W$-gates that we have to perform 
to implement our proposal for $n$ qudits is $\frac{n(n-1)(2n-1)}{6}$,
the number of single qubit and CNOT gates required by our circuit scales as
$O\left( n^3 (\log_2 n)^2 \right)$.

There's a second way of encoding a qudit in the state $\ket{i}$, that is,
take $n$ qubits, set the $i$-th one to $\ket{1}$, 
and then the rest to $\ket{0}$,
i. e. $\ket{i}=\ket{0\ldots 010 \ldots 0}$.
This unary encoding is less efficient than the previous one,
since it requires $n$ qubits compared to the $\log_2 n$ required before.
Nevertheless, it has the advantage that allows us to implement
the $W$-gates using less gates.
We just follow the previous three steps with the four qubits that have to 
be affected by the gate, that is, the $i$-th, $j$-th, $(n+i)$-th and $(n+j)$-th. 
We realize that only 3 different multi controlled gates are required. In this case, then,
the number of single qubit and CNOT gates required by our circuit would be
$O\left( n^3 \right)$.

%\section{General comments}
Moreover, let us note that we can transform
our antisymmetrization circuit into a symmetrization one
by just changing the signs of the definition of the $W$ gates
in Eq. (\ref{eq:W-gates}), i.e. 
$\sqrt{p}\ket{ij} \mp \sqrt{1-p}\ket{ji} \ \to \ \sqrt{p}\ket{ij} \pm \sqrt{1-p}\ket{ji}$.
Another possibility of performing the same symmetrization would be 
to invert the order of the input state of the circuit ($\ket{n-1,\ldots, 0}$
instead of $\ket{0,\ldots, n-1}$) and to apply the gates $W_{n-1-i,n-1-j}$, instead of
$W_{ij}$, along the circuit.
In both cases, the W gates  always act in the positive combination and
the final state obtained is fully symmetric in all possible permutations.

Let us conclude with some comments on the generalization of our
proposal to other values of $m$.
The first observation is that, if $m > 1$, the number of states
that appear in the superposition of the Laughlin wave function
is much larger than the simple permutations of the input single states. 
The corresponding quantum circuit, therefore, 
cannot be only composed of $W$-gates. 
This will increase a lot the degrees of freedom of our
elementary gates and, thus, its complexity. Though specific examples
for low values of $n$ and $m$ can be found, a general scheme
is still missing.

%\begin{acknowledgements}
{\bf Acknowledgements}. Financial support from QAP (EU), ICINN (Spain), 
FI program and Grup Consolidat  (Generalitat de Catalunya), 
and QOIT Consolider-Ingenio 2010 is acknowledged.

%\end{acknowledgements}


\begin{thebibliography}{10}

\bibitem{F82}   R. Feynman, Internat. J. Theoret. Phys., 21, pp 467-488 (1982).

\bibitem{NC00} M. A. Nielsen, and I. L. Chuang, {\it Quantum Computation and Quantum Information}, 
               (Cambridge Univ. Press, Cambridge, England, 2000).

% REVIEWS DE OPTICAL LATTICES PER QUANTUM SIMULATORS
%\bibitem{LSADSS07}M. Lewenstein, A. Sanpera, V. Ahufinger, B. Damski, A. Sen De and U. Sen, 
%Advances in Physics Vol. 56 Nos. 1-2, 243-379 (2007).

\bibitem{LSADSS07-BDZ08}M. Lewenstein et al. , Adv. in Phys.  56, 243 (2007). 
                          %{\it Ultracold atomic gases in optical lattices: mimicking condensed matter physics and beyond}
                           I. Bloch, J. Dalibard and W. Zwerger, Rev. Mod. Phys. 80, 885 (2008).
                          %{\it Many-Body Physics with Ultracold Gases}



% PROPOSTES DE QUANTUM SIMULATORS D'ESTATS TOPOLŽÃ'GICS I LAUGHLIN
\bibitem{PPC} M. Popp, B. Paredes and J. I. Cirac, Phys. Rev. A 70, 053612 (2004).
 %{\it Adiabatic Path to Fractional Quantum Hall States of a Few Bosonic Atoms}

%\bibitem{HSDL07} M. Hafezi, A. S. Sorensen, E. Demler and M. D. Lukin, Phys. Rev. A 76, 023613 (2007).
\bibitem{HSDL07} M. Hafezi et al., Phys. Rev. A 76, 023613 (2007).
%{\it Fractional Quantum Hall Effect in optical lattices}

\bibitem{BHHFZ05} H. P. B\"{u}chler et al. ,
                  % M. Hermele, S. D. Huber, M. P. Fisher and P. Zoller,
                  Phys. Rev. Lett. {\bf 95}, 040402 (2005).
%{\it Atomic quantum simulator for lattice gauge theories and ring exchange models}

\bibitem{PB08} B. Paredes and I. Bloch, {\it cond-mat/0711.3796}, (2007).

\bibitem{MBZ06} A. Micheli, G.K. Brennen and P. Zoller, Nat. Phys. 2, 341 (2006).
%{\it A toolbox for lattice-spin models with polar molecules}

\bibitem{DDL03} L. M. Duan, E. Demler and M. D. Lukin, Phys. Rev. Lett. {\bf 91} 090402 (2003).
%{\it Controlling Spin Exchange Interactions of Ultracold Atoms in Optical Lattices}


% SIMULACIO DE MQ AMB UN ORDINADOR CLASSIC
\bibitem{G03} G. Vidal, Phys. Rev. Lett. {\bf 91} 147902 (2003). 
              G. Vidal, Phys. Rev. Lett. {\bf 93} 040502 (2004).


% PROPOSTES CIRCUITS QUANTICS
\bibitem{BS08} G. Benenti and G. Strini, Am. J. Phys. 76  657 (2008).

\bibitem{VCL08} F. Vestraete, J. I. Cirac and J. I. Latorre, {\it quant-ph/0804.1888}, (2008).

% LAUGHLIN STATE AND FQHE
\bibitem{L83} R. B. Laughlin, Phys. Rev. Lett. {\bf 50}, 1395 (1983).

\bibitem{Tsui82} D. C. Tsui, H. L. Stormer and A. C. Gossard, Phys. Rev. Lett. {\bf 48} 1559 (1982).

\bibitem{ILO07} S. Iblisdir, J. I. Latorre and R. Or\'us, Phys. Rev. Lett. {\bf 98} 060402 (2007).

% CNOT, SINGLE GATES AND UNIVERSAL SET 
\bibitem{MMKIW95} C. Monroe et al. ,
                  %D. M. Meekhof, B. E. King, W. M. Itano and D. J. Wineland,
                  Phys. Rev. Lett. {\bf 75}, 4714 (1995).

\bibitem{BBCDMSSSW95} A. Barenco et al. ,
                      %C. H. Bennett, R. Cleve, D. P. DiVincenzo, N. Margolus, 
                      %P. Shor, T. Sleator,   J. A. Smolin and H. Weinfurter,
                      Phys. Rev. A 52, 3457 (1995).


% REALITZACIO EXPERIMENTAL DE QUANTUM COMPUTERS
\bibitem{CZ95} J. I. Cirac and P. Zoller, Phys. Rev. Lett. {\bf 74}, 4091 (1995).
%  {\it CNOT amb trapped ions}

\bibitem{CCEHRSZ06} G. Chen et al. , 
   %D. Church, B. Englert,. C. Henkel, B. Rohwedder, M. Scully and. M. Zubairy,
  {\it Quantum Computing Devices: Principles, Designs, and Analysis}, (CRC Press, 2006).


\end{thebibliography}
\end{document}